\documentclass[11pt]{article}
\usepackage{graphics}

\begin{document}
\title{\textsf{\textbf{Strengthening of a Polymer Interface:\\ Interdiffusion and 
Crosslinking}}}
\author{A.\ Aradian\footnote{Achod.Aradian@college-de-france.fr}, E.\ Rapha\"{e}l
\footnote{Elie.Raphael@college-de-france.fr}, and P.-G.\ de Gennes
\footnote{Pierre-Gilles.deGennes@espci.fr} \\
Coll\`{e}ge de France\footnote{Laboratoire de Physique de la Mati\`ere Condens\'ee,
C.N.R.S.\ URA 792 }, 11 place Marcelin Berthelot \\ 75231 Paris Cedex
05, France}
\maketitle           
\begin{abstract}
In many industrial processes, pieces of the same polymer material are brought
into contact at a temperature above the glass transition. Interdiffusion 
takes place across the interface and leads to a strengthening
of the junction. Often, a crosslinker agent is also added in order to improve the
global mechanical properties of the material, as in 
the formation of latex films from dispersed solutions of polymer particles. We studied 
theoretically the competition between the interdiffusion and the crosslinking reaction, and found that 
the control parameter tuning the balance between these two processes is $\alpha = Q \tau_0 A_0^* N^3 b^3/N_e$, 
where $Q \tau_0$ accounts for the reactivity of the crosslinker, $A_0^*$ 
is the initial concentration of sites capable of crosslinking on 
the polymer chains, $N$ is the polymerization index, $N_e$ the number of segments between entanglements
and $b$ a distance comparable to the segment length. The case of practical 
interest is $\alpha \ll 1$: the reaction locks the interfacial chains 
once a significant mixing has developed, resulting in films with good
mechanical properties.  
\end{abstract}        
\section{Introduction}
When two pieces of the same polymer material are brought into good contact at a temperature
above their glass transition temperature, the macroscopic interface between 
the pieces progressively disappears, whereas the mechanical strength of 
the interface increases. This phenomenon, of great practical importance, 
is known as ``polymer-polymer welding'' or ``crack healing'',
and has been widely studied, both experimentally~\cite{Woolbook} and 
theoretically~\cite{PGGCras1980,PGGSanchez,Prager,Wool}. The crack healing is primarily due
to the diffusion of the polymer chains from both sides across the interface.

In many situations, a crosslinker agent is added into the material: this is 
notably the case in the technology of latex coatings, where individual latex particles from
a polymer dispersion are cast onto a surface (see the review on the subject by M.A.\ Winnik 
in ref~\cite{Winnikreview}). Upon drying, particles form contacts and 
progressively coalesce to give a continuous film, whose mechanical
properties may be significantly improved by crosslinker addition~\cite{TaylorWinnik,Bufkin}.

The use of a crosslinker, however, brings some difficulties: a fine 
balance between interdiffusion and crosslinking rates is required
to obtain optimal film strength, because reaction and diffusion 
enter into competition. The interdiffusion is strongly sensitive 
to molecular weight, and slows down when the crosslinking 
reaction advances and the chains become more and more branched. Thus if 
the reaction rate is too fast, particles
will mix only partially, to the disadvantage of film 
tenacity. As this issue is crucial to industrial applications, experimental studies
have been carried out monitoring the evolution of the interface at a microscopic
level~\cite{Feng,Tamai}.

Our aim in the present article, is to try, from a simple 
analysis based on scaling laws, to extract the important parameters that 
control the final state of the interface, and attempt to find some 
guidelines in optimizing systems displaying both interdiffusion and crosslinking.
We do not hope, given the complexity of real situations,
to afford a set of quantitative predictions.
    
The article is organized as follows. In section 2, we present our 
model, and derive the governing equations of the problem. 
In section 3, which will be of main interest for practical 
applications, we exhibit a control parameter and present the 
results of the model. In section 4, after discussing some limitations 
of our approach, we try to relate a macroscopic quantity as the 
adhesion energy to our previous microscopic results, and finally, 
we briefly consider a few other systems used in film coating 
technology that take benefit of alternative strengthening strategies.
\section{Governing equations}
\subsection{Assumptions of the model}
Initially, at the onset of contact between polymer particles, 
the chains near the surface are ``reflected'' at the interfaces (see Figure 1).
\begin{figure}
\centering
\resizebox{0.5 \textwidth}{!}{%
\includegraphics*[5.5cm,13.5cm][15cm,22.4cm]{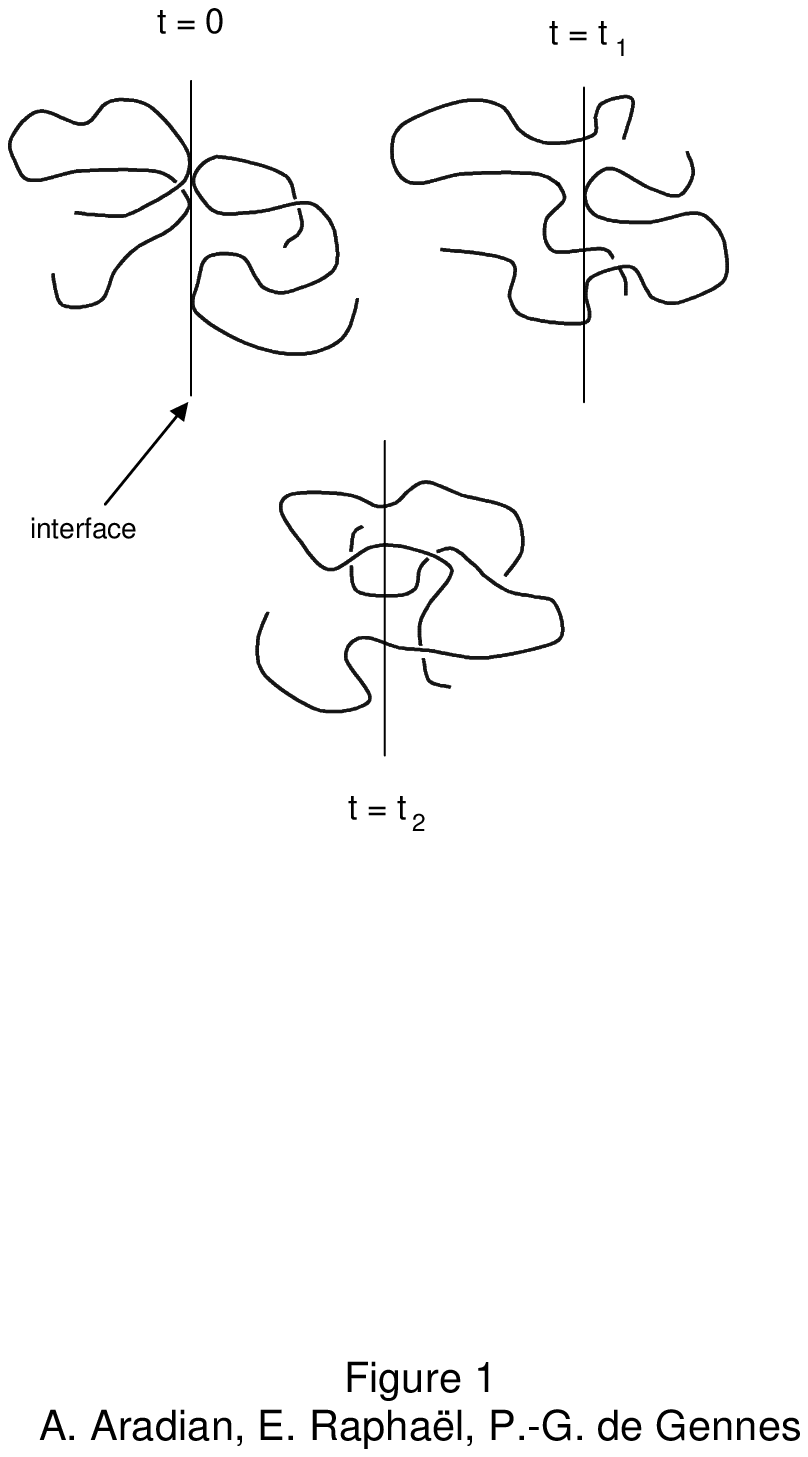}
}  
\caption{Aspect of interfacial chains at successive instants. 
Initially ($t=0$), the chains are reflected at the interface, then chain ends 
start exploring the new volume across the border ($t=t_1$), so that conformations
progressively relax to the Gaussian, equilibrium shape ($t=t_2$).}
\end{figure}
Because of the 
new volume made available to them after contact, these distorted conformations progressively relax 
towards the equilibrium, Gaussian conformations of chains in a melt, which maximize the entropy of 
the system. However, as shown in~\cite{PGGCras1980,PGGSanchez}, 
this process does not take place through a Fickian 
diffusion of the chains: because of topological entanglements, the chains are trapped in the
reflected tubes, and the only way to explore the volume on the opposite side of the interface 
is through the motion of chain ends which open the path across the junction and progressively 
drag along the rest of the chain according to a reptation mechanism.

In our system, a crosslinking agent is also added 
into the polymer particles, prior to contact. Hence, a chemical reaction proceeds 
in parallel to chain diffusion, and forms bridges between chains. This, in turn, has 
large consequences on the diffusion process itself, because 
crosslinked chains form branched objects whose motions are exponentially slower than linear
chains~\cite{PGGStar}: interdiffusion dramatically slows down, as well as the reaction rate (which is 
related to chain motion). In the general case, providing a description 
of the subsequent evolution of the system is extremely complex, 
with a vast variety of macromolecules reacting one 
with the other (all with different numbers, positions and lengths of 
branches, and hence, different diffusion kinetics). Our 
description will thus rely on a few simplifying assumptions, which 
we are now going to present, deferring a critical discussion to 
section 4.

We consider two latex particles of the same polymer, which are brought 
into contact at time $t=0$. The temperature is assumed to stay constant, and to exceed
the glass transition temperature of the samples. The macromolecules, which we assume to be 
linear and monodisperse, consist of $N$ units ($N$ taken greater than the entanglement 
threshold $N_e$), and are statistical copolymers of two types of monomers, 
$A$ and $A^*$. The crosslinker, called $X$, is a bi-functional agent, able to 
bind only to $A^*$ sites. The crosslinking reaction 
sets up in two steps. First, the crosslinker molecules $X$ attach to the ``active'' 
sites $A^*$, yielding $PA^*+X \to$ $PA^*$--$X$ (where $P$ stands for ``polymer backbone''). 
Then, in a second reaction, the ``true'' crosslinking between 
chains occurs, yielding $PA^*$--$X + A^*P \to PA^*$--$X$--$A^*P$. The first step
involves the diffusion of a small molecule $X$ and should be very 
fast compared to the second step, which involves reaction of sites both borne by 
polymer chains. For this reason, in the rest of the article, we will consider only the 
last step (polymer-polymer reaction), assuming that the attachment step is entirely 
completed at $t=0$~\cite{initialcross}.

We also restrict ourselves to the case of only \emph{one} $X$ attached per 
chain. This assumption is not meant solely to bring a simplification, but is also 
founded on practical grounds: from the classical work of Gent and coworkers~\cite{Gent1,Gent2}, 
we know that 
for elastomers, a tighter network, as resulting from an increased $X$ concentration,
makes a poorer adhesive. Our hope is thus that this hypothesis still allows 
for most situations of interest.

Finally, we assume that when a chain binds to another, the resulting branched object
remains \emph{fixed} in position; this is in contrast with \emph{mobile} chains, 
that did not undergo any reaction (except the binding of an $X$ molecule) and 
still keep their full mobility (see Figure 2). 
\begin{figure}
\centering
\includegraphics*[7.5cm,15.3cm][13.2cm,21.1cm]{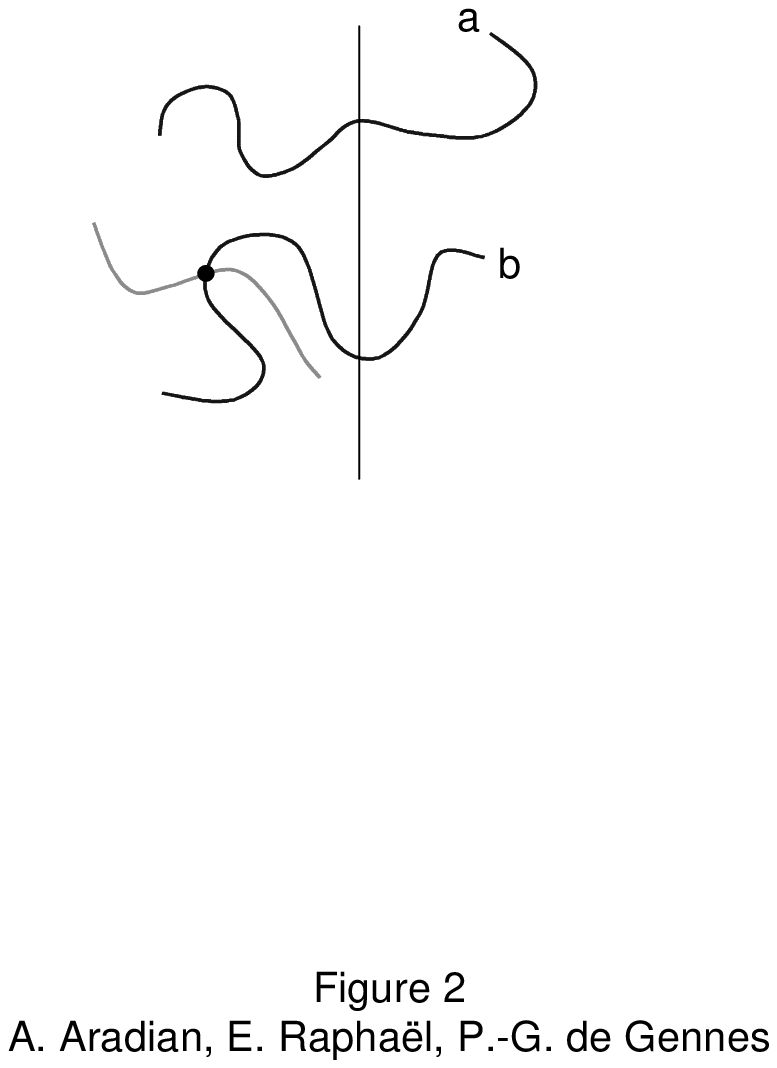}
\caption{Different types of chains: Chain `a' is mobile; chain `b' is considered
to be fixed, because it formed a crosslink (black dot) with another (grey) chain 
(the vertical line depicts the initial interface between the adjacent particles).}
\end{figure}
Of course, this assumption is not 
entirely realistic, because branched objects have a finite -though 
reduced- mobility. We shall come back to this point at the end of 
the article.
\subsection{Extent of mixing $\Gamma_f$}
We add crosslinkers to enhance the mechanical strength of the overall
film that is obtained after coalescence of the particles. At the microscopic 
scale of chains, the strengthening of an interface is due to the interdiffusion between the two 
blocks, {\it i.e.}, the extent of mixing between adjacent 
particles. At a given time, if we consider the chains that have crossed the interface, 
some are fixed (because of prior crosslinking) so that their contribution to the 
mixing is permanent, and some are mobile and their 
contribution is bound to evolve. As the crosslinking reaction proceeds, the
mobile chains ``disappear'' to the benefit of fixed ones, so that sooner or later, 
any chain that once contributed in a ``transient'' way  to the extent of mixing
finally contributes ``permanently''. Ultimately, all chains
are fixed and the extent of mixing reaches a maximum value which characterizes the final state
of the film. In the following, because it is of greater physical importance, we will 
focus our attention mainly on the permanent part of the extent of mixing, $\Gamma_f$ 
(taking only fixed chains into account), and will consider the transient part $\Gamma_m$
 only incidentally.

As stated already, the motion of the chains is actually led by the motion 
of chain ends across the interface. Let us define $\rho_0$ as the 
initial ($t=0$) density (per unit volume) of mobile chain ends at position $x$, 
and make the hypothesis that it is uniform (no segregation of chain ends 
at the interface). Denoting $x$ the abscissa on the axis perpendicular to
the interface (which is located at $x=0$), we also define $\rho_m (x,t)$ 
and $\rho_f (x,t)$ as, respectively, the densities of mobile chain ends and fixed
chain ends at position $x$ and time $t$. Our first relation comes from the conservation 
of the total number of chain ends, yielding:
\begin{equation}
\label{conservation}
\frac{d\rho_m}{dt}(x,t)=-\frac{d\rho_f}{dt}(x,t).
\end{equation}

The next point is to notice that, as far as the ongoing crosslinking reaction is concerned, 
the medium is spatially homogeneous. The argument proceeds as follows. At $t=0$, 
chains near the interface indeed start from out-of-equilibrium 
configurations. But, to first order, if we assume that the contact is perfect 
and that the melt density is the same at the interface and in the 
bulk, the diffusion modes of these distorted chains are the same 
as those of equilibrium bulk chains. As it is intimately related 
to these diffusion processes, the reaction rate must accordingly remain 
uniform in the system. In other words, this means 
that, wherever it is located, any given unit volume of the system hosts 
the same number of reactions.

In order to keep track of the reaction advancement, let us call $r(t)$ 
the `reacted fraction', {\it i.e.}, the fraction of the initially mobile 
chains whose $A^*X$ site has reacted between time $t=0$ and time $t$ (implying $r=0$ at 
$t=0$, and $r=1$ at the end of the reaction). Each time a chain reacts, it becomes a 
fixed branched object, so that $r(t)$ also gives the fraction of mobile chains that 
have become fixed by reaction, yielding~\cite{stopped}:
\begin{equation}\label{rhomobile}
\rho_m (x,t)=\rho_0[1-r(t)].
\end{equation}
From this, using eq~\ref{conservation}, we deduce  
\begin{equation}
\label{rhofixed}
\rho_f(x,t)=-\int_0^t \frac{d\rho_m}{dt} dt=\rho_f(t=0) + \rho_0 
r(t),
\end{equation}                                         
where $\rho_f(t=0)$ is the initial density of fixed chains (initial network).

Let us describe the evolution of the interdiffusion for times smaller than 
the reptation time of the chains $T_{rep}$. At $t=0$, the film is made of the juxtaposition
of polyhedral cells~\cite{Winnikreview}, displaying flat interfaces with their neighbors 
(Figure 3).
\begin{figure}
\centering
\includegraphics*[5.5cm,15.5cm][15cm,19.4cm]{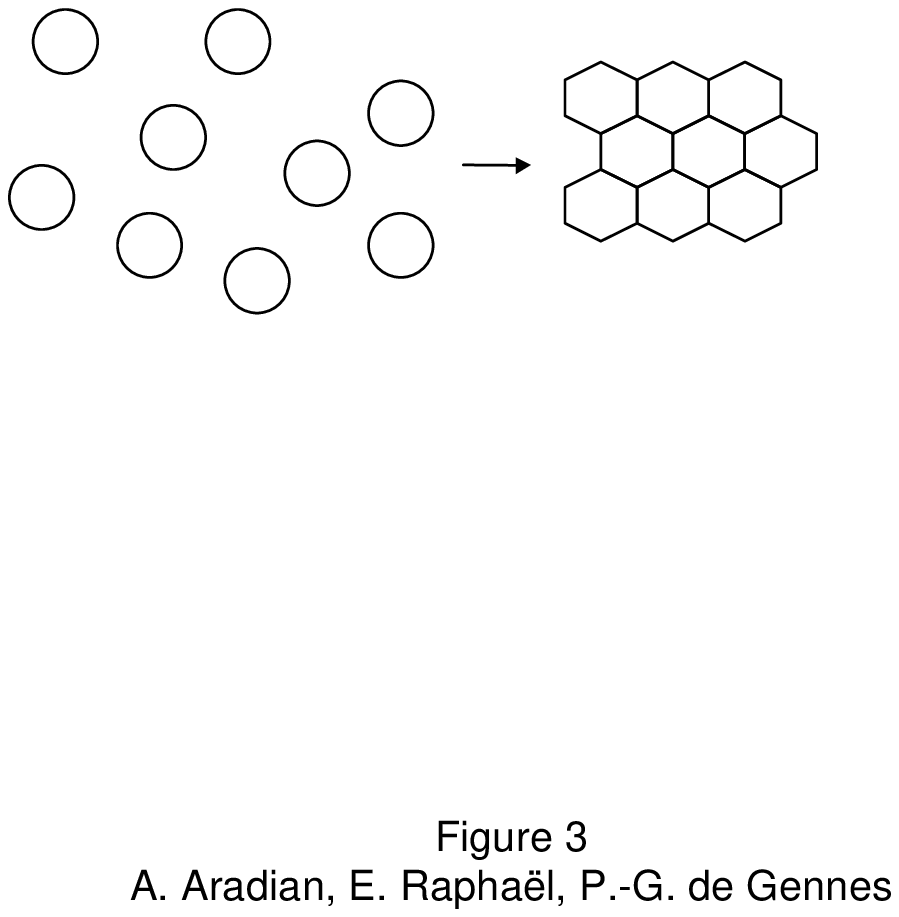}
\caption{When an aqueous dispersion of spherical polymer particles dries (on the left), 
the particles enter into contact and are deformed into space-filling polyhedral 
cells (on the right). This is the initial $t=0$ situation of our model, where we 
consider only such flat interfaces. Neighboring cells then progressively coalesce 
to form a continuous film.}
\end{figure}
Adjacent cells then start coalescing by interdiffusion. In a given cell 1, 
we consider a thin slice of material, chosen parallel to the interface
 with cell 2, and located at some position $x$ (see Figure 4). 
\begin{figure}
\centering
\includegraphics*[5.8cm,14.8cm][13.7cm,21.5cm]{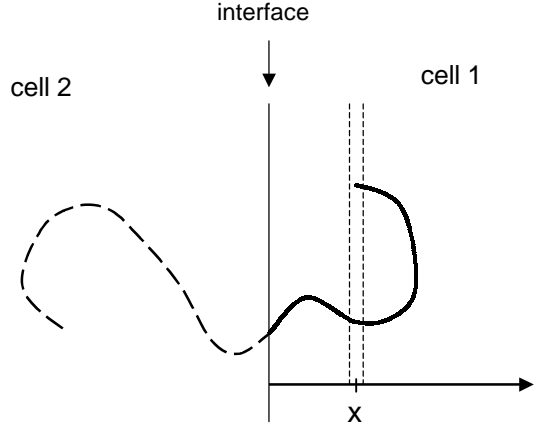}
\caption{A chain from cell 2 invading cell 1, with its end located in a
slice at a distance $x$ from the interface. The number of monomers in the ``invading
portion'' of the chain (solid line) can be estimated to $\simeq x^2/a^2$. The monomers 
in the dashed part of the chain should not enter into the computation of  
the extent of mixing.}
\end{figure}
When 
do the first chain ends coming from cell 2 start to invade this slice? To answer this 
question, we have to know the distance traveled by a chain end in a given time $t$.
According to the reptation theory~\cite{PGGreptation,DoiEdwards}, each chain diffuses 
inside a tube, and over a time $t$, 
travels a (r.m.s.) tube length given by $s(t)= a (N/\sqrt{N_e}) (t/T_{rep})^{1/2}$, where 
$a$ denotes the length of a chain segment, and $T_{rep}=N^3\tau_0/N_e$ is the reptation 
time of the chain ($\tau_0$ the relaxation time of a monomer). To this curvilinear length corresponds
a distance ``as the crow flies'', $l_{rep}(t)$, which is given by:
\begin{equation}
\label{defofl}
l_{rep}(t)= (s a \sqrt{N_e})^{1/2}=R_0 \left( \frac{t}{T_{rep}} \right)^{1/4} \mbox{\qquad} 
(t \: \raisebox{-.4ex}{$\stackrel{<}{\sim}$} \: T_{rep}),
\end{equation}
where $R_0=a N^{1/2}$ is the Gaussian extension of a chain. Within constants, 
this distance $l_{rep}(t)$ is also the (rectilinear) distance traveled by the 
chain ends. Inverting this formula, we can now see that the first 
chain ends from cell 2 reach the slice at $x$ in cell 1 at a time
\begin{equation}
\label{defoftx}
t_x=T_{rep} \left( \frac{x}{R_0} \right)^4.
\end{equation}                                                      

We now want to compute $\Gamma_f$, the extent of mixing due to permanent chains. 
All chains passing through the slice do not pursue their 
migration, a fraction of them is stopped there by crosslinking. This 
situation is accounted for by the local augmentation of $\rho_f$. From $t=0$, 
the density of \emph{crossing} chain ends $\rho_f^{cross}$ (from cell 2) that 
have settled at $x$ (in cell 1) may then be written as
\begin{equation} \label{rhoinv}
\rho_f^{cross}(x,t) = \: \left\{
\begin{array}{ll}
\rho_f(t)-\rho_f(t_x) & \mbox{\qquad if } t>t_x \\
0 & \mbox{\qquad if } t \leq t_x
\end{array}
\right.
\end{equation}
(where we took into account that chain ends settling down at times earlier than $t_x$
do not come from cell 2). Of course, eq~\ref{rhoinv} is not exact, and oversimplifies the 
density profiles generated by the diffusion process, but it should retain the main features.

Counting that each invading chain end located at $x$ drags along approximately 
$x^2/a^2$ monomers, because dragged chains form random walks (see Figure 4), and 
summing over all possible slices, 
we are now able to give a formula for the extent of mixing $\Gamma_f$:
\begin{equation}
\label{gammaf}
\Gamma_f(t)=\int_0^{\infty}\rho_f^{cross}(x,t) \frac{x^2}{a^2}dx \mbox{\qquad} 
(t \: \raisebox{-.4ex}{$\stackrel{<}{\sim}$} \: T_{rep}).
\end{equation}
Finally, with the use of $\rho_f^{cross}$ (eq~\ref{rhoinv}) and $\rho_f$ 
(eq~\ref{rhofixed}), we arrive at
\begin{equation}\label{gammaf2}
\Gamma_f(t)= \frac{\rho_0}{a^2} \left( 
r(t)l_{rep}^3(t)-\int_0^{l_{rep}(t)}r(t_x)x^2dx \right) \mbox{\qquad} 
(t \: \raisebox{-.4ex}{$\stackrel{<}{\sim}$} \: T_{rep})
\end{equation}
(ignoring numerical factors, which would not be very meaningful at our scaling law level).

We can also calculate the contribution $\Gamma_m$ from mobile 
chains. In this case, with obvious notations, we have $\rho_m^{cross}(x,t)= \rho_m(t)-\rho_m(t_x)$ if 
$t>t_x$, and $\rho_m^{cross}(x,t)= 0$ otherwise, and thus
\begin{equation}\label{gammam}
\Gamma_m(t)=\int_0^{\infty}\rho_m^{cross}(x,t) \frac{x^2}{a^2}dx=\frac{\rho_0}{a^2}[1-r(t)]
l_{rep}^3(t) \mbox{\qquad} (t \: \raisebox{-.4ex}{$\stackrel{<}{\sim}$} \: T_{rep}).    
\end{equation}

We now turn to the evolution of the system after the reptation 
time $T_{rep}$. Chains from each side of the interface keep 
diffusing further into the cell on the other side, so that the 
mixing between adjacent cells keeps increasing. However, for
time $t > T_{rep}$, the traveled distance $l_{rep}(t)$ becomes 
greater than the Gaussian radius of the chains $R_0$. Thus, some 
chains migrate farther than a distance $R_0$ from the interface  
and lose any intersection with it, so that they do not bridge the 
junction anymore: although they improve the mixing between cells, 
such chains do not contribute to the strength of the interface. 
Since our interest lies in keeping track of the strengthening of the 
interface, rather than of the mixing by itself, we must discard these 
leaving chains from our computation, and keep only the ``efficient'' 
part of $\Gamma_f$ and $\Gamma_m$ (see Figure 5).
\begin{figure}
\centering
\includegraphics*[7cm,14cm][14.3cm,19.5cm]{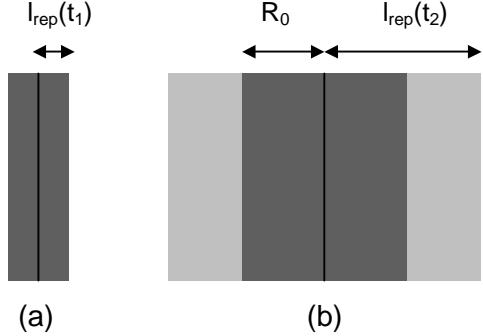}
\caption{Extent of mixing between two cells, at two different times $t_1$ and $t_2$ 
(the vertical, black, line depicts the interface). (a) At time $t_1 < T_{rep}$, 
the mixing occurs inside the shaded region of breadth $l_{rep}(t_1)$.
(b) At time $t_2 > T_{rep}$, the mixing has increased and takes place over a region 
of width $l_{rep}(t_2)$ (dark and light regions). However, only the chains located 
inside the dark region of width $R_0$ effectively bridge the interface: from the 
point of view of interface strength, this is the ``efficient'' part of the mixing.}
\end{figure}
This is done by saturating the length $l_{rep}$ to $R_0$ after time 
$T_{rep}$ in equations \ref{gammaf2}-\ref{gammam}. The modified 
equations at times greater than $T_{rep}$ take the form: 
 
\begin{equation}
\label{gammafsat}
\Gamma_f(t)= \frac{\rho_0 R_0^3}{a^2}r(t)
- \frac{\rho_0 R_0^3}{a^2} \int_0^{R_0}r(t_x)x^2dx \mbox{\qquad} 
(t \: \raisebox{-.4ex}{$\stackrel{>}{\sim}$} \: T_{rep}),
\end{equation}                                                          
and
\begin{equation}\label{gammamsat}
\Gamma_m(t)=\frac{\rho_0 R_0^3}{a^2}[1-r(t)]
\mbox{\qquad} (t \: \raisebox{-.4ex}{$\stackrel{>}{\sim}$} \: T_{rep}).    
\end{equation}
We notice that, in eq~\ref{gammafsat}, the second term containing the 
integral is now a constant independent of time. 
\subsection{Reacted fraction r(t)}
In section 2.2, the reacted fraction $r(t)$ was defined as the fraction of the mobile 
chains present initially whose $A^*X$ site has reacted with 
another site between $t=0$ and time $t$. Initially, at $t=0$, all the $A^*X$ sites 
are unreacted, yielding $r=0$. At the end of the reaction, all the $A^*X$ 
have reacted and $r=1$. To complete our calculation of $\Gamma_f$ and 
$\Gamma_m$, we now need to compute the reacted fraction $r(t)$.

Because of the non-Fickian laws for segmental diffusion in 
polymeric materials, reaction kinetics may display very special 
features in the so-called `diffusion-controlled regime'~\cite{PGGCompact}.
 However, for reagents of common reactivity, the 
practical case is usually the `mean-field regime'~\cite{OShauLetter}, that is to 
say, the regime of classical, small molecules, reaction kinetics, 
and this is indeed the regime that will be assumed in the following.

We start by defining $Q$ as the `local' reactivity of the active 
sites $A^*X$, to be understood as the probability of reaction per 
unit time of a pair $A^*X/A^*$ provided they are in permanent 
contact. Then the reaction constant $k$ 
is~\cite{PGGCompact,OShauMacro}
\begin{equation}
k=Qb^3,
\end{equation}
where $b$ is the capture radius, {\it i.e.}, the distance below 
which reaction becomes possible. This capture radius is a molecular
length, and we will often simply assume $b \simeq a$ (chain unit 
size).

We define $X_0$ as the initial ($t=0$) concentration of $A^*$ 
sites bound to an $X$ molecule, {\it i.e.}, the initial concentration in $A^*X$ groups. 
$A^*_0$ denotes the initial concentration in the \emph{remaining} $A^*$ sites, {\it i.e.}, 
not bound to any $X$. Initially, the concentration of intermolecular bridges 
$A^*XA^*$ is zero. We also define a parameter $\eta=X_0/A^*_0$. 
In our assumptions, there is only one $X$ attached per chain, but $A^*$ sites may be more 
numerous, so that we necessarily have $\eta \leq 1$. A straightforward calculation of 
chemical kinetics yields the number of $A^*X$ sites that have been consumed by the bridging 
reaction $A^*X+A^* \to A^*XA^*$. Then, to compute the desired fraction $r(t)$ from this number,
we simply need to divide it by the initial concentration $X_0$.  

For $\eta <1$, we obtain
\begin{equation}
r(t)=\frac{1-\exp[ -(1-\eta)A_0^*Qb^3t]}{1-\eta 
\exp [-(1-\eta)A_0^*Qb^3t]} \mbox{\qquad} (\eta<1),
\end{equation}
which can be easily rearranged to introduce the reptation time 
$T_{rep}=N^3\tau_0/N_e$ into 
\begin{equation}\label{rinf1}
r(t)=\frac{1-\exp[ -(1-\eta) Q \tau_0 \mu \; (t/T_{rep})]}{1-\eta 
\exp[-(1-\eta)Q \tau_0 \mu \; (t/T_{rep})]} \mbox{\qquad} 
(\eta<1),
\end{equation}
where a new important quantity appears: $\mu \equiv A_0^*N^3b^3/N_e$. 

For the marginal case $\eta=1$, a similar calculation can be 
performed, and yields
\begin{equation}\label{req1}
r(t)=\frac{Q \tau_0 \mu \; (t/T_{rep})}{1 + Q \tau_0 \mu \; (t/T_{rep})}
\mbox{\qquad} (\eta =1).
\end{equation}
\section{Results}
We have seen in the previous sections, that when two particles come into
contact, the interface strengthens because of the interdiffusion  
and the crosslinking of chains at the junction. At a microscopic level, 
the extent of mixing between the adjacent particles $\Gamma_f$, 
due to fixed (crosslinked) chains, provides a good account of this strengthening. 
We are now able to give results for this extent of mixing, using 
eqs~\ref{gammaf2}-\ref{gammamsat} for the extent of mixing and 
eqs~\ref{rinf1}-\ref{req1} for the reacted fraction $r(t)$.

We will split our discussion into two cases, depending on the relative rates 
of the crosslinking reaction and of the interdiffusion process. As we shall see, 
the crucial control parameter in the system is
\begin{equation}\label{alpha}
\alpha=Q \tau_0 \mu = Q\tau_0 A_0^* \frac{N^3}{N_e} b^3
\end{equation}
In eq~\ref{alpha}, $Q \tau_0$ is a dimensionless number, corresponding 
to the probability of reaction for \emph{one} collision of reactive sites 
($Q$ is the probability per unit time, 
and $\tau_0$, the monomer relaxation time, is similar to the collision duration).
\subsection{`Fast reaction' regime: $\alpha \gg 1$}
We start with the case $\alpha \gg1$. If we consider the reaction rate r(t)
 (eq~\ref{rinf1}), we easily see that 
$r(t) \simeq 1$ is reached (end of reaction) for $t \gg 
\frac{T_{rep}}{(1-\eta) \, \alpha}$, {\it i.e.}, at times much smaller than 
$T_{rep}$ since $\alpha \gg 1$. (We stress that here $\eta$ 
cannot be chosen too close to 1, or we must use another expression for $r(t)$).
Evaluating $\Gamma_f$ (eq~\ref{gammaf2}) at times great enough so that the 
reaction can safely be considered as complete, we find that the final 
value is:
\begin{equation}\label{gammafinalsmall}
\Gamma_f^{final} \simeq \frac{\rho_0 R_0^3}{a^2} \: 
\frac{1}{[(1-\eta)\alpha]^{3/4}}
\end{equation}
(within prefactors of order unity), which, as will be seen shortly,
is a  very small value compared to the maximum reachable extent of mixing 
$\Gamma_f^{max} \simeq \rho_0 R_0^3 / a^2$.

We should also mention the marginal case, when $\eta$ is chosen 
very close to 1. Then the correct expression for $r(t)$ is given 
by eq~\ref{req1}, but $\Gamma_f^{final}$ remains very similar (and 
very small):
\begin{equation}
\label{gammaffast}
\Gamma_f^{final} \simeq \frac{\rho_0 R_0^3}{a^2} \: 
\frac{1}{\alpha^{3/4}}
\end{equation}
(within prefactors).

We conclude that this situation is not favorable to the development of a 
strong junction at the particles' interfaces, and should be avoided. Actually, 
in this case, the reaction is so fast compared to the interdiffusion, that all
chains are fixed by crosslinking well before they could bridge the interface
efficiently.  
\subsection{`Slow reaction' regime: $\alpha \ll 1$}
We now take $\alpha \ll 1$. Here the reaction is complete at 
times $t \gg T_{rep}/\alpha$ 
which are larger than $T_{rep}$. To evaluate $\Gamma_f^{final}$, 
we shall then use the form valid at times greater than $T_{rep}$, 
as given by eq~\ref{gammafsat}. In this expression, one can 
actually neglect the contribution from the integral, with respect 
to the first term, because for $x \leq R_0$, $r(t_x) \ll 1$. The 
extent of mixing $\Gamma_f$ hence evolves as $r(t) \: \rho_0 
R_0^3/a^2$, and reaches a final value
\begin{equation}
\label{gammafslow}
\Gamma_f^{final} \simeq \frac{\rho_0 R_0^3}{a^2},
\end{equation}
which is the largest physically realizable value in our 
system.

We note that, if we have $\eta \simeq 1$, using the appropriate 
expression of $r(t)$ yields the same result.

This `slow reaction' regime is thus the interesting one for 
practical purposes, because one allows the interdiffusion to 
develop fully before locking the formed bridges in position. We 
may also add that, in practice, $\alpha$ will probably not need 
to be much smaller than unity for the system to display a good 
strengthening.

One might wonder whether $\alpha \ll 1$ is an easy value to find 
in real systems. If we exclude highly reactive groups, like 
radicals, usually $Q \tau_0$ is well below $10^{-6}$~\cite{OShauEPJE}.
The concentration of $A^*$ sites also plays an important role. If 
there are a few $A^*$ sites per chain, $A^*_0 \simeq 1/Na^3$, then 
$\alpha \simeq Q \tau_0 (N^2/N_e) (b/a)^3$, which is around $10^{-4}$ with 
$Q \tau_0=10^{-8}$, $N=10^3$, $N_e=10^2$ and $b \simeq a$. With many $A^*$ on 
each chain, say $A_0^* \simeq p/a^3$ ($p<1$ but not too small), 
we get $\alpha \simeq Q \tau_0 (N^3/N_e) p$, which can still be made small:
$Q \tau_0=10^{-8}$, $N=10^3$, $N_e=10^2$, $p=0.1$, yield $\alpha = 0.01$.
\subsection{A simple interpretation of $\alpha$}
We will here provide a simple explanation of the control 
parameter $\alpha=Q \tau_0 A_0^*N^3b^3/N_e$ that has been exhibited above. 
For this purpose, let us consider an $A^*X$ group located on a 
given chain chosen at random among the others, and follow its 
evolution from the beginning of the crosslinking reaction, at 
$t=0$.

Thinking in terms of a lattice model, where the $A^*X$ 
molecule moves with a jump frequency $1/\tau_0$ (where $\tau_0$ is 
the monomer relaxation time), we may say that the $A^*X$ group makes 
$T_{rep}/\tau_0$ displacements from $t=0$ until $t=T_{rep}$. At 
each of these movements, the $A^*X$ molecule may react and form a 
crosslink with any of the $A^*$ molecules lying within a capture 
volume $b^3$. In our mean-field situation, we can estimate the 
number of such candidates approximately as $A_0^*b^3$. However, 
only a fraction of these collisions with the possible candidates 
bring an effective reaction: with $Q$ the `local' reactivity, 
{\it i.e.}, the probability of effective reaction per unit time when 
an $A^*X$ encounters an $A^*$, and $\tau_0$ the contact time, we find 
that an $A^*X$ molecule may have $Q \tau_0 A_0^*b^3$ effective 
crosslinking reactions at each displacement. (Note that this is only a virtual
number of reactions, since an $A^*X$ site is able to react only
once.) Hence, over a time $T_{rep}$, we deduce that an $A^*X$ group (virtually) makes
$Q \tau_0 A_0^*b^3 T_{rep}/\tau_0$ reactions. Noticing that $T_{rep}=N^3\tau_0/N_e$, 
we see that this quantity is the same as $\alpha$. Thus, $\alpha$ 
can be understood as the number of crosslinks that a given chain may form by 
reaction of its $A^*X$ site, if it were able to react an infinite number of times.

We can now go further and understand the fast and slow reaction 
regimes in terms of characteristic times. On one hand, we define 
a characteristic crosslinking time as the time required for an 
$A^*X$ site to react (once) with an $A^*$ molecule. Since an 
$A^*X$ site may have $\alpha$ (virtual) reactions in a time 
$T_{rep}$, we estimate this characteristic time for one reaction as 
$T_{rep}/\alpha$. On the other hand, the characteristic time for 
the interdiffusion process is $T_{rep}$, {\it i.e.}, the time 
needed for an interfacial chain to relax entirely from its initial 
contorted conformation.

Comparing these characteristic times together, we are able to retrieve,
in a simple way, some conclusions that have been exposed previously. When 
$\alpha \gg 1$ ({\it i.e.}, $T_{rep}/\alpha \ll T_{rep}$), the reaction is 
faster than the interdiffusion, locking the chains before they can 
bridge the interface. On the opposite, when $\alpha \ll 1$ ({\it i.e.}, 
$T_{rep}/\alpha \gg T_{rep}$), the reaction mainly occurs after an 
appreciable amount of interdiffusion has set up, and enhances the 
mechanical properties of the film.                                                  
\section{Discussion}
\subsection{Further remarks}
The approach that has been presented here is rather simplified, 
and could be improved in several ways. We list below a few points
that would need to be worked on in order to get a more accurate description
of the evolution of the system, that would go farther than scaling laws.

i) We restricted ourselves to monodisperse systems, but practical situations
may prove more difficult. One of the effects of polydispersity is to give rise
to a range of chain diffusion coefficients, and hence to a blurring of simple scaling
laws as the ones that were used here (this has already been observed on systems displaying 
interdiffusion alone~\cite{WangWinnik}). Other effects of polydispersity are discussed
in Taylor and Winnik~\cite{TaylorWinnik}. Another point to notice is 
that, in real systems, having \emph{precisely} one $X$ molecule bound per 
polymer chain is unfeasible: the number of $X$ per chain will rather display a 
statistical distribution of possible values, with an average of order unity. This distribution may
also present spatial variations inside each particle: in the example of latex
films, before contact between individual particles is set, the $X$ crosslinker molecules are
originally outside the particles, in the aqueous solution, and progressively diffuse inside. 
The concentration profile arising from such a diffusion process will inevitably present spatial 
dependences.

ii) As pointed out in our model assumptions, we consider that branched objects
remain fixed. This is not completely true, and one may imagine that in the late stages
of the reaction most unreacted sites are borne by arms belonging to branched molecules.
A complete theory would thus certainly need to take the kinetics resulting from reactions
between such molecules into account. Very slow evolutions were indeed 
observed in experiments on certain types of latex 
films~\cite{PhamWinnik}.

iii) Another assumption of our approach is that the kinetics of 
polymer reactions remains the same as that of small molecules, due to the weak reactivity of
usual chemical species. However, as was shown by O'Shaughnessy in~\cite{OShauMacro},
there is necessarily a crossover at long times, from the usual kinetics regime to the 
`diffusion-controlled' (DC) regime resulting from the non-Fickian diffusion of segments
(the so-called `compact exploration'). We checked that, with usual chain lengths
and usual reactivities of crosslinking sites, the crossover would take place well after
completion of the crosslinking reaction. Would the DC regime set up earlier, for example 
if we choose very reactive agents ($Q \tau_0 \simeq 1$), this would be harmful to the 
mechanical properties of the interface, and thus an undesirable situation.

iv) A last question is related to the surface state of the particles before they
start merging together. In order to stabilize the initial solution, surface-active agents 
may be present. After onset of contacts, they might modify the diffusion kinetics of the chains across 
the interfaces. Also, with certain polymer compositions, the particles themselves may have a 
`core-shell' structure, where the outer part of the particles display distinct properties
compared to the inside.  
\subsection{Estimation of the adhesion energy}
From a macroscopic point of view, the strengthening of the
interfaces may be characterized by several physical quantities.
One of them is the adhesion energy, defined as the work required
to separate two particles by opening a fracture at the interface.
We will here present an attempt to estimate this energy, in the limit 
of very low separation velocities.

We start with the fast reaction case $\alpha \gg1$ (section 3.1),
and consider the final state, after completion of the reaction.
Most of the chains were stopped by reaction very early in the process
of interdiffusion. If we look at chains originally from the left side
of the interface, we notice that these reactions most probably
occurred on the longest part of the chain, {\it i.e.}, the portion still
in the original particle on the left. If we now try to separate the two
adjacent particles, we will have to pull out the portions of chains
that have penetrated into the right-hand particle (see Figure 6).
\begin{figure}
\centering
\includegraphics*[6.5cm,14.4cm][13.5cm,21.3cm]{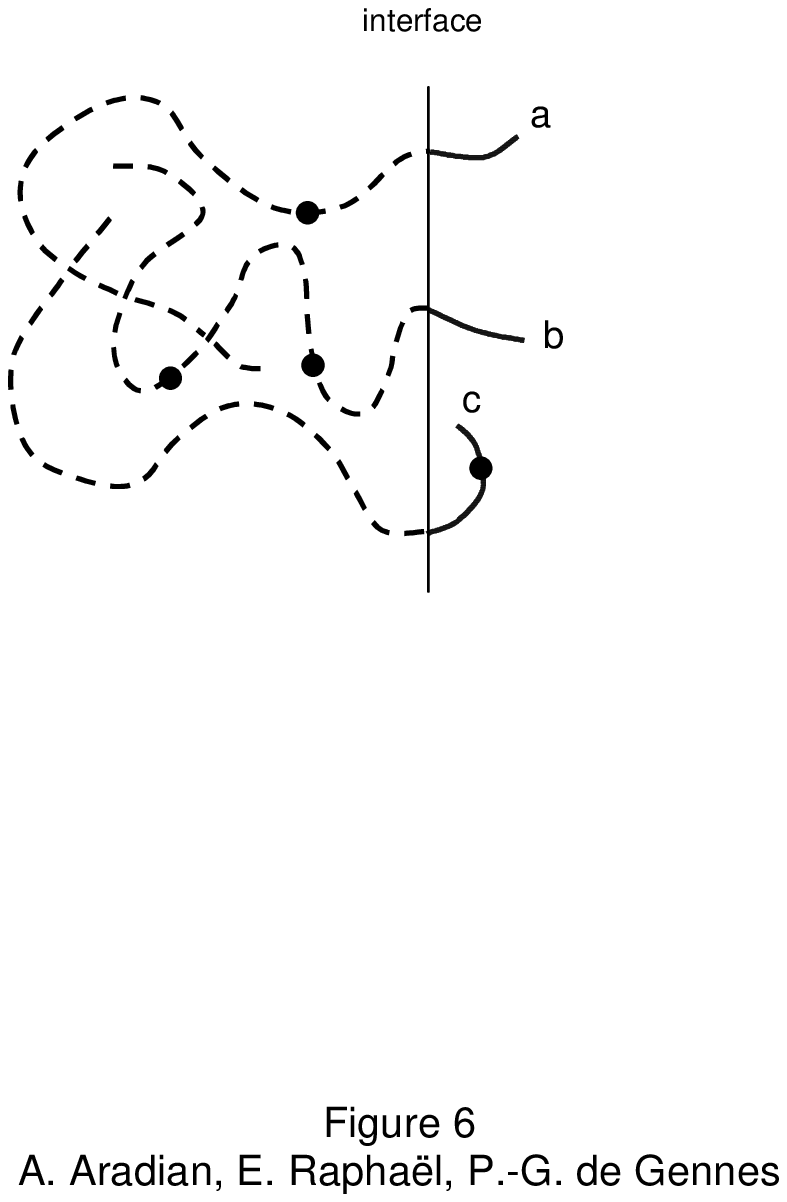}
\caption{Chains at the interface in the final state of the fast reaction regime.
Chains `a' and `b' depict the most common case, where the crosslinks (black dots)
occurred on the portions which remained in their home particle (dotted lines). 
For chain `c', the reaction occurred in the portion inside the other particle, 
but this a rare event (shortest arm). When a fracture opens, most chains
(like `a' and `b') will pull the portions in solid lines out.}
\end{figure}
Some years ago, Rapha\"el and de Gennes developed a 
model~\cite{EliePGG} evaluating the adhesion energy when 
one pulls `connector' chains grafted on a surface out of an elastomer. 
It was found that the adhesion energy $G$ in the quasi-static limit (very slow
separation) is given by
\begin{equation}\label{adhElie}
G=U_{v} \sigma n,
\end{equation}
where $U_{v}$ is a Van der Waals energy, $\sigma$ the number of 
connectors per unit area of interface, and $n$ the polymerization 
index of the connectors. If we try to extend this expression to our 
situation, we must integrate eq~\ref{adhElie} over a distribution 
of connector lengths, since the effective pull-out length that has 
to be counted is the (varying) portion inserted into the adjacent 
particle (and not the total chain length $N$). Using the notations 
of sec.\ 2.2, it is easy to find that the appropriate form is
\begin{equation}
G=U_{v}\: \int_0^{\infty}\rho_f^{cross}(x,t) \frac{x^2}{a^2}dx
\end{equation}
(where the time $t$ in the integral is chosen after completion of 
the reaction). This expression appears to be similar to eq~\ref{gammaf}
for the extent of mixing. Thus, we conclude that in this case 
($\alpha \gg 1$), the zero-rate adhesion energy $G$ would be 
simply proportional to the final extent of mixing:
\begin{equation} \label{Gfast}
G \simeq U_{v} \Gamma_f^{final}
\end{equation}
(with $\Gamma_f^{final}$ given by eq~\ref{gammafinalsmall} or~\ref{gammaffast}).

We now consider the slow reaction regime ($\alpha \ll 1$). In this case, as already stated,
the interface heals before the reaction has significantly advanced, so that 
most interfacial chains spread arms of comparable size ($\sim N$) on both sides 
of the junction by the time they are pinned down. On Figure 7, which presents some possible
configurations of the interfacial chains, we see that in addition to the chains that will have
to be pulled out if we try to open a fracture, some chains are 
anchored on both sides and will rather undergo a scission. (The contribution of such scission    
processes to the adhesion energy $G$ is estimated in the Appendix.) 
\begin{figure}
\centering
\includegraphics*[7cm,14.8cm][13.7cm,20cm]{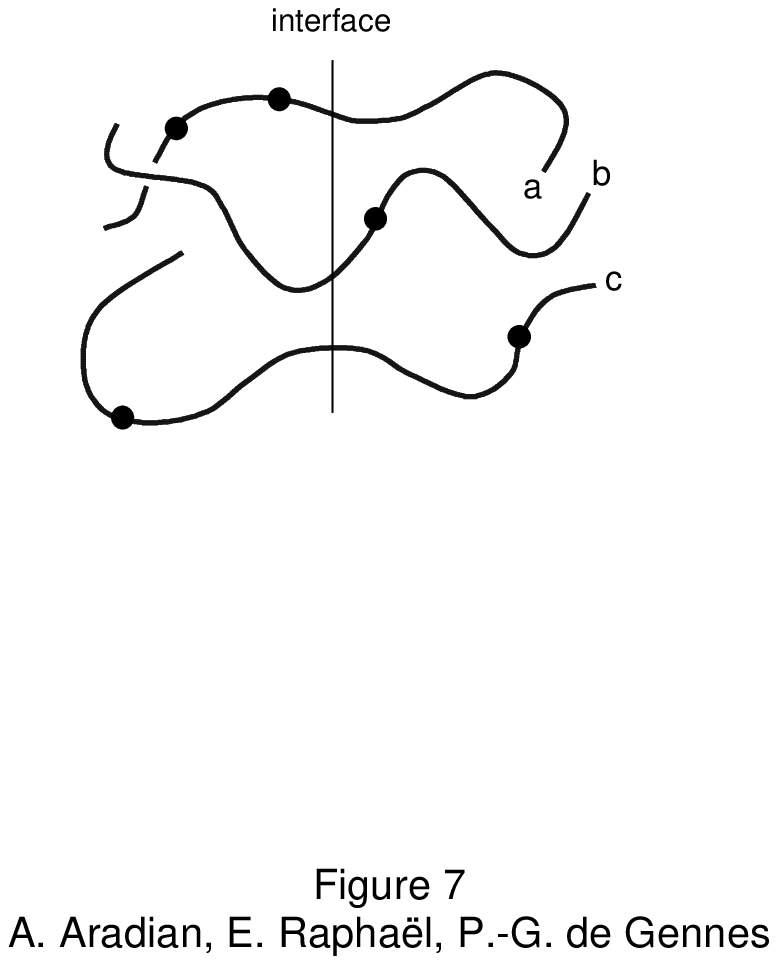}
\caption{Chains at the interface in the final state of the slow reaction regime.
Most chains have arms of comparable size on both sides of the 
interface. Upon opening of a fracture, chain `a' will pull its right-side 
portion out, chain `b' will pull its left-side portion out, and chain `c' will
undergo a scission process.}
\end{figure}
If we assume that the scission
and the pull-out contribution simply add to each other, we find that, again, $G$ is 
related to the final extent of mixing $\Gamma_f^{final}$ by a simple proportionality, 
in a similar way to eq~\ref{Gfast}:
\begin{equation} \label{Gslow}
G \simeq U_{0} \Gamma_f^{final},
\end{equation}
where $U_{0}$ is an energy, usually comparable to that of a chemical bond, and $\Gamma_f^{final}$ 
should be taken as in eq~\ref{gammafslow}). A derivation of 
this result is available in the Appendix.

Let us now consider the molecular weight dependence of these 
(zero-rate) adhesion energies in the fast and slow reaction 
regimes. In the \emph{fast reaction} regime, we have $G \simeq U_{v} 
\Gamma_f^{final}$ (eq~\ref{Gfast}) with $\Gamma_f^{final}$ as 
given by eq~\ref{gammafinalsmall} or~\ref{gammaffast}. The dependence
in $N$ of the parameter $\alpha=Q \tau_0 A_0^* N^3 b^3/N_e$ can be evaluated 
as follows: in situations where the initial concentration $A_0^*$ is a 
finite, $N$-independent, fraction of the total monomer concentration ($A_0^* 
= p/a^3$, $p<1$), we have $\alpha \sim N^3$. Alternately, if there are only 
a few $A^*$ sites on each chain ($A_0^* \simeq 1/Na^3$), we have $\alpha \sim N^2$.
Together with $\rho_0 \sim 1/N$, $R_0 \sim N^{1/2}$, we get
\begin{equation} \label{GfastN}
G_{fast} \sim \: \left\{
\begin{array}{ll}
N^{-7/4} & (A_0^*= p/a^3) \\
N^{-1} &   (A_0^* \simeq 1/Na^3).
\end{array}
\right.
\end{equation}
On the other hand, it is easy to estimate the $N$-dependence of the 
\emph{slow reaction} regime, by using $G \simeq U_{0} \Gamma_f^{final}$ 
(eq~\ref{Gslow}) and $\Gamma_f^{final} \simeq \rho_0 R_0^3/a^2$                 
(eq~\ref{gammafslow}):
\begin{equation} \label{GslowN}
G_{slow} \sim N^{1/2}
\end{equation}
It appears that the fast and slow reaction regimes exhibit clearly 
distinct dependences on the molecular weight of the polymer: $G_{fast}$
has a marked inverse dependence on $N$, whereas $G_{slow}$ displays a 
square-root, increasing, dependence. Such predictions may prove testable 
on the experimental side, either by direct macroscopic measurements of the 
adhesion strength, or by microscopic techniques surveying the local evolution 
of the interpenetration (one would then rather have access to the extent of 
mixing $\Gamma_f^{final}$, but molecular weight dependences are identical).      

In our approach, the adhesion energy $G$ is proportional to the final extent 
of mixing $\Gamma_f^{final}$ (obtained after completion of the 
crosslinking reaction). Since $\Gamma_f^{final}$ saturates when the 
interpenetration distance is of the order of the Gaussian size $R_0$ 
of the chains (see section 2.2), $G$ saturates to its maximum value at the same time. 
This result is related to the fact that our systems are maintained above the glass
transition (allowing relative sliding of chains), and that 
we are dealing with adhesion energies at low separation velocities. The property that 
the full tensile strength of a latex film is established over 
a time comparable to, or larger than, the reptation time is supported by 
several experimental studies (see ref.~\cite{Winnikreview} and references therein). 
We emphasize, however, that glassy polymers may display rather different 
features~\cite{glassy}.

We conclude this section by stressing that all these estimations of the 
adhesion work remain of course very crude, but they suggest 
a simple link between a microscopic quantity like the extent of mixing and the 
macroscopic adhesion energy.       
\subsection{Other systems}
In the previous sections, we considered a system where the strenghtening 
of the interface occurs in the presence of a crosslinking agent. When the system
is optimized, chains have time to diffuse across the interfaces before being locked
by the chemical reaction and form permanent bridges. This strategy is not the only 
available, and we will here briefly consider a few other approaches of 
use in the latex film technology.

1) One possibility is to use a copolymer containing as one of the comonomers
a functionality $Y$ which, upon addition of a catalyst, is able to react with another 
$Y$ to form a dimer. This reaction leads to the formation of $Y$--$Y$ bridges between 
polymer chains, throughout the system, at the interfaces and in the bulk of the 
polymeric material. We can see that this is very similar to our system with a 
crosslinker agent, as soon as we neglect the first attachment of 
the crosslinker to the macromolecules.

2) A second possibility is to use a mixture of two different kinds of 
particles, for instance one sort made of a polymer A, and the other
made of a polymer B, where chains of polymer A bear sites able to 
react with sites of polymer B. We may further assume that 
polymer A and B are strongly immiscible ($\chi>0$ and $\chi N \gg 1$, with $\chi$ the 
associated Flory parameter), so that the region where A and B 
coexist is very thin compared to the chain radius. In this case, A 
and B chains react only within the interface, and form block 
copolymers at the junction (see Figure 8). 
\begin{figure}
\centering
\includegraphics*[7.8cm,15cm][14.2cm,21.5cm]{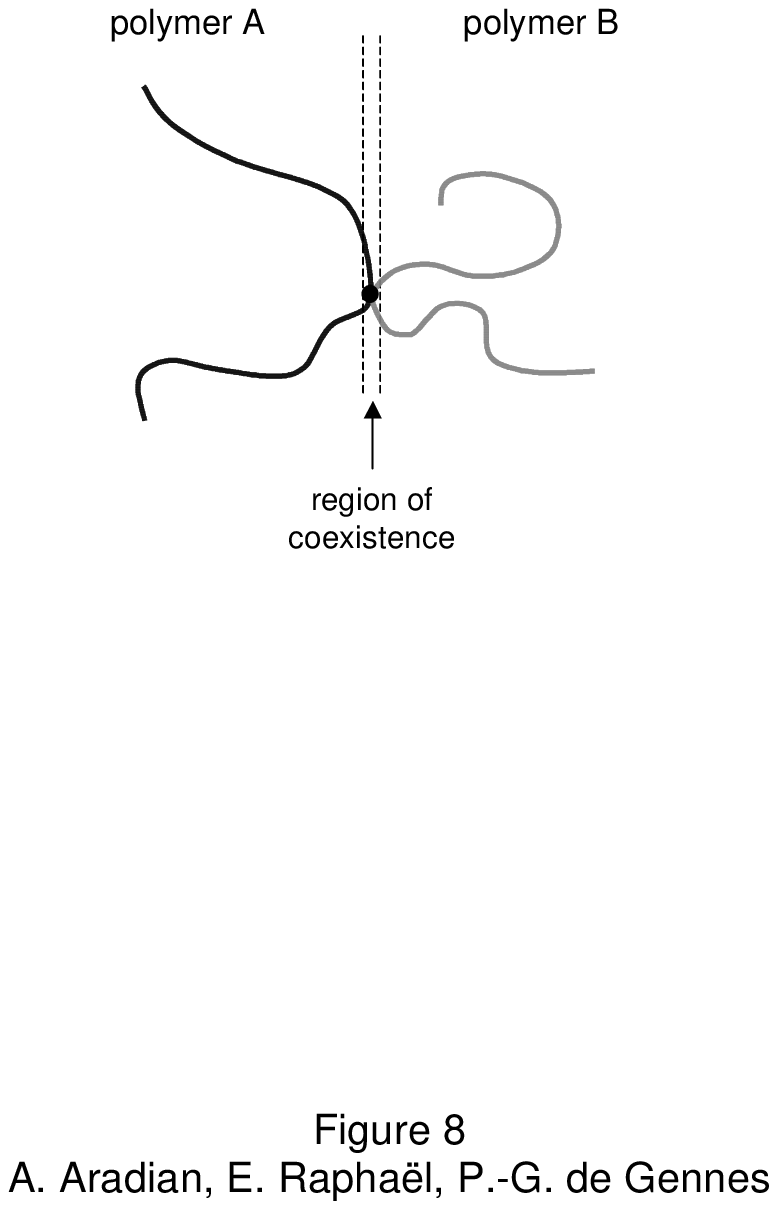}
\caption{Block copolymer formed by reaction at the border region between 
two incompatible polymers. Chains from each kind remain confined to their
own particle, but the formation of crosslinks in the coexistence region 
creates copolymer bridges that will have to be pulled out if a fracture opens.}
\end{figure}
Theory actually predicts a host of 
different chemical kinetics regimes for the formation of 
such bridges~\cite{Fredrickson,OShauinterf1,OShauinterf2}. There is, 
in the present situation, a significant strengthening of the interface, 
even though the extent of mixing remains vanishingly small: as seen on Figure 8,
here again, one must pull out the interfacial copolymers out
of their surrounding matrix to open a fracture. To characterize the strength of the
interface, the adhesion energy formula of eq~\ref{adhElie} discussed above seems 
suitable, and we may use it to give a few guidelines to 
optimize A/B systems. From eq~\ref{adhElie}, we see that we need the longest 
possible attached strands at the interface: this would suggest to 
have only \emph{one} reactive site per chain (for A as well as for 
B). We would also wish to reach the highest possible interface 
coverage by connectors ($\sigma \simeq \rho_0 R_0$), a condition 
which should be easy to fulfill, because it is a 
natural tendency of the interfacial reaction~\cite{Fredrickson,OShauinterf1,OShauinterf2}.
Then the interfacial adhesion energy would be of 
order
\begin{equation}
G \simeq U_0 \: \frac{1}{a^2 \: N^{1/2}} \: N \simeq G_{max},
\end{equation}
similar to the value reached with crosslinked systems.

3) Another variant of the A/B system is the case of a \emph{miscible} pair 
A/B. This situation, however, raises difficulties 
of its own: reactions occur in a non-stationary, spreading region, making the 
computation of reactional quantities very difficult. In the case of 
small molecules with a time-dependent reaction front, theoretical 
studies~\cite{Racz,Lee} predict anomalous laws for reaction rates. 
But, to our best knowledge, the macromolecule case (of interest to us) still 
remains to be clarified.
\section*{Acknowledgements}
It is a pleasure to thank Prof.\ M.A.\ Winnik of the University of 
Toronto, who introduced us to this subject, and helped us forward 
with fruitful exchanges. The authors are also thankful to C.\ 
Creton, L.\ L\'eger, and the anonymous referees, for 
discussions and useful comments.
\section*{Appendix. Estimation of the adhesion energy in the slow reaction regime}
In this Appendix, we want to estimate the adhesion energy between the particles in 
the final state of the slow reaction regime (Figure 7). 

We start by evaluating the density of connectors chains (crossing chains) 
$\sigma_f$ at the end of the reaction. If we consider an area S 
of interface, the chains crossing it must lie 
within a distance $R_0$, so that there is a number $\rho_0 R_0 S$ of them.
This corresponds to a grafting density $\sigma_f \simeq \rho_0 R_0$.

A fraction $f_1$ of these crossing chains have reacted only on one side of the interface,
and thus contribute to the adhesion by a pull-out mechanism. Considering that most 
crossing chains display arms of comparable size $\sim N$ on both sides of the 
interface, and using eq~\ref{adhElie} to compute the corresponding adhesion energy, 
we find a contribution $G_{pull-out} \simeq U_{v} f_1 \sigma_f N$.
 
But a fraction $f_2$ of the interfacial chains is anchored 
on \emph{both} sides of the interface (Figure 7), and rather than being pulled out, 
they undergo a scission when the interfacial fracture opens. 
For this scission mechanism, we may evaluate 
the adhesion energy using the classical Lake and Thomas argument~\cite{LakeThomas}: 
at the moment of rupture \emph{each} monomer along one connector chain has stored
an energy comparable to the chemical bond energy $U_\chi$. Hence, to bring a chain
to rupture, we have to provide an energy $U_\chi$ to all the monomers between the 
anchorage points. Since crosslinked points are not too numerous on each chain, 
in average the number of monomers under tension is $\sim N$ (within factors). We 
may then evaluate the scission contribution to the adhesion energy as 
$G_{scission} \simeq U_\chi f_2 \sigma_f N$.

We note that $G_{pullout}$ and $G_{scission}$ have 
the same structure (at this scaling law level). If we assume that 
these energies add to each other into a global adhesion energy $G_{tot}$,
we have $G_{tot} \simeq U_0 \sigma_f N$, 
where the energy $U_0$ would be a weighted average of $U_\chi$ and 
$U_{v}$ with respect to the probabilities of the scission and 
pull-out mechanisms. Then, using $\sigma_f=\rho_0 R_0$ and $N=R_0^2/a^2$,
we see that this adhesion energy is related to the final extent of mixing 
(eq~\ref{gammafslow}) in a proportional way:
\begin{equation}
G_{tot} \simeq U_0 \: \rho_0 R_0 \: R_0^2/a^2 \simeq U_0 
\Gamma_f^{final},
\end{equation}
as stated in section 4.2.
We add that, because the covalent energy $U_\chi$ is much 
larger than the van der Waals energy $U_v$ ($U_\chi/U_v \simeq 40$),
$U_0$ should be, in many cases, dominated by the scission contribution, 
or equivalently, $G_{tot} \simeq G_{scission}$. 


\begin{thebibliography}{99}
\bibitem{initialcross}
To be more specific, in the context of filmification from a
dispersion of latex particles in water, the crosslinker agent 
is originally in the aqueous solution, and diffuses inside the 
polymer particles. We suppose in the text that this diffusion inside 
the polymer phases and the subsequent attachment reaction (our so-called
``first step'') are fast processes (compared to the polymer reactions),
so that they are completed during the evaporation of the solvent, or very quickly
after onset of contacts. Even further, in many real cases, not only the first step 
is finished at the time of contact, but the polymer-polymer reaction step 
has already begun inside individual particles, in the dispersed state. 
Such a premature crosslinking reduces the amount of mobile chains that will be 
available for interdiffusion when the particles come in contact (at $t=0$).
\end{thebibliography}
\end{document}